\documentclass[aps,prx,twocolumn,footinbib,superscriptaddress,showpacs,preprintnumbers]{revtex4-2}

\usepackage[utf8]{inputenc}

\usepackage{cancel}
\usepackage{comment}

\usepackage[dvipsnames]{xcolor}      % use if color is used in text
\definecolor{lcolor}{rgb}{0.5,0,0}
\definecolor{citcolor}{rgb}{0,0.0,1}

\usepackage[breaklinks,colorlinks,urlcolor=blue,citecolor=citcolor,linkcolor=lcolor,linktoc=all]{hyperref}
\usepackage{color}
\usepackage{graphicx}	% Include figure field
\graphicspath{{./figures/}}
\usepackage[utf8]{inputenc}
\usepackage{amsmath}
\usepackage{amssymb}
\usepackage{amsfonts}
\usepackage[ragged]{footmisc}
\usepackage{slashed}
\usepackage{float}
\usepackage[capitalise]{cleveref}
\usepackage{dsfont}
\usepackage{enumerate}

\usepackage[normalem]{ulem}

\usepackage{mathrsfs}
\usepackage{tikz}
\usepackage[com pat=1.1.0]{tikz-feynman}
\usetikzlibrary{calc}
\usepackage{xcolor}
\usepackage{slashed}
\usepackage{MnSymbol}
\usepackage{dsfont}
\usepackage{mathtools}
\usepackage{wrapfig}
\usepackage{hyperref}
\usepackage{bbold}
\usepackage{comment}
\usepackage{MnSymbol} %for \sumint
\usepackage{array}

\makeatletter
\g@addto@macro\bfseries{\boldmath}
\makeatother

    \DeclareFontFamily{U}{wncy}{}
    \DeclareFontShape{U}{wncy}{m}{n}{<->wncyr10}{}
    \DeclareSymbolFont{mcy}{U}{wncy}{m}{n}
    \DeclareMathSymbol{\Sh}{\mathord}{mcy}{"58}

\usepackage{tikz}
\usepackage[customcolors]{hf-tikz}

\usetikzlibrary{arrows,cd,shapes,decorations.pathmorphing,decorations.markings,shadings}
\tikzset{
  big arrow/.style={
    decoration={markings,mark=at position 1 with {\arrow[scale=4,#1]{>}}},
    postaction={decorate},
    shorten >=0.4pt},
  big arrow/.default=blue}

\newcommand{\bbone}{\text{\usefont{U}{bbold}{m}{n}1}}
\MakeRobust{\bbone}
\usepackage[acronym]{glossaries}
\newacronym{LO}{LO}{leading-order}

\newacronym{QM}{QM}{quark matter}

\newacronym{QCD}{QCD}{quantum chromodynamics}

\newacronym{HTL}{HTL}{Hard-Thermal-Loop}

\newacronym{UV}{UV}{ultraviolet}

\newacronym{IR}{IR}{infrared}

\newcommand{\MSbar}{$\overline{\text{MS}}$}

\begin{document}

\title{Complete $g^6$ pressure of hot Yang--Mills theory}

\preprint{HIP-2026-20/TH}

\author{Keijo Kajantie}
\email{keijo.kajantie@helsinki.fi}
\affiliation{Department of Physics and Helsinki Institute of Physics, P.O.~Box 64, FI-00014 University of Helsinki, Finland}
\author{Pablo Navarrete}
\email{pablo.navarrete@helsinki.fi}
\affiliation{Department of Physics and Helsinki Institute of Physics, P.O.~Box 64, FI-00014 University of Helsinki, Finland}
\author{Risto Paatelainen}
\email{risto.paatelainen@utu.fi}
\affiliation{Department of Physics and Astronomy, FI-20014 University of Turku, Finland}
\affiliation{Department of Physics and Helsinki Institute of Physics, P.O.~Box 64, FI-00014 University of Helsinki, Finland}
\author{\\Leon Sandbote}
\email{leon.sandbote@helsinki.fi}
\affiliation{Department of Physics and Helsinki Institute of Physics, P.O.~Box 64, FI-00014 University of Helsinki, Finland}
\author{York Schröder}
\email{yschroder@ubiobio.cl}
\affiliation{Centro de Ciencias Exactas, Departamento de Ciencias Básicas, Universidad del Bío-Bío, Avenida
Andrés Bello 720, Chillán, Chile}
\author{Kaapo Seppänen}
\email{kaapo.seppaenen@unibe.ch}
\affiliation{AEC, Institute for Theoretical Physics, University of Bern, Sidlerstrasse 5, CH-3012 Bern, Switzerland}
\author{Aleksi Vuorinen}
\email{aleksi.vuorinen@helsinki.fi}
\affiliation{Department of Physics and Helsinki Institute of Physics, P.O.~Box 64, FI-00014 University of Helsinki, Finland}

\begin{abstract}

We present the full order $g^6$ term in the weak-coupling expansion of the pressure of hot $\mathrm{SU}(3)$ Yang-Mills theory, resolving a problem that has remained open for more than two decades.
To achieve this, we evaluate nine previously unknown nonfactorizing four-loop sum-integrals using the thermal loop-tree duality method with novel local ultraviolet and infrared subtractions, thereby completing the hard-scale contribution to the pressure for general $\mathrm{SU}(N_c)$.
Combining this result with known chromoelectric and chromomagnetic contributions, available for $N_c=3$, we demonstrate the analytic cancellation of infrared divergences originating from different momentum scales and obtain the complete $O(g^6)$ pressure for $\mathrm{SU}(3)$.
Our result displays quantitative agreement with data from four-dimensional lattice simulations over a broad range of temperatures, providing a stringent test of the dimensionally reduced effective-field-theory paradigm.

\end{abstract}

\maketitle

\section{Introduction}

Pure $\mathrm{SU}(N_c)$ Yang-Mills (YM) theory provides a very favorable setting for studying the thermodynamics of strongly interacting quantum field theories at high temperature.
Despite the absence of fermionic fields, it retains the central features of quantum chromodynamics (QCD), including asymptotic freedom \cite{Gross:1973id,Politzer:1973fx} and confinement, with a deconfinement transition that is second order for $N_c=2$ and first order for $N_c\geq3$ \cite{Svetitsky:1982gs,Lucini:2003zr,Lucini:2005vg,Panero:2009tv}.
Its relative computational simplicity allows precision lattice calculations, making it a valuable testing ground for both perturbative and nonperturbative approaches to thermal field theory \cite{Andersen:2009tc,Borsanyi:2012ve,Giusti:2014ila}.

At sufficiently high temperatures $T$, where we may assume the gauge coupling to satisfy $g\ll1$, Yang-Mills plasma exhibits three parametrically separated momentum scales: the hard scale $2\pi T$, associated with nonzero Matsubara frequencies; the soft scale $gT$, associated with electrostatic screening; and the ultrasoft scale $g^2T$, associated with nonperturbative magnetostatic dynamics.
Dimensional reduction \cite{Appelquist:1981vg,Braaten:1995cm,Kajantie:1995dw} exploits this hierarchy by successively integrating out the hard and soft modes.
The resulting three-dimensional effective field theories (EFTs), electrostatic QCD (EQCD) and magnetostatic QCD (MQCD), describe the static electric and magnetic sectors and the ultrasoft magnetic sector, respectively. 
They contribute to the Yang-Mills pressure via \cite{Braaten:1995jr} 
\begin{equation}
p_{\rm YM}=p_{\rm E}+p_{\rm M}+p_{\rm G}\, ,
\label{eq:completepressure}
\end{equation}
where the three terms stand for the naive loop expansion of the full-theory pressure, the perturbatively evaluated EQCD pressure, and the nonperturbatively determined MQCD pressure.

\begin{figure*}[t]
\centering
\includegraphics[width=0.95\textwidth]{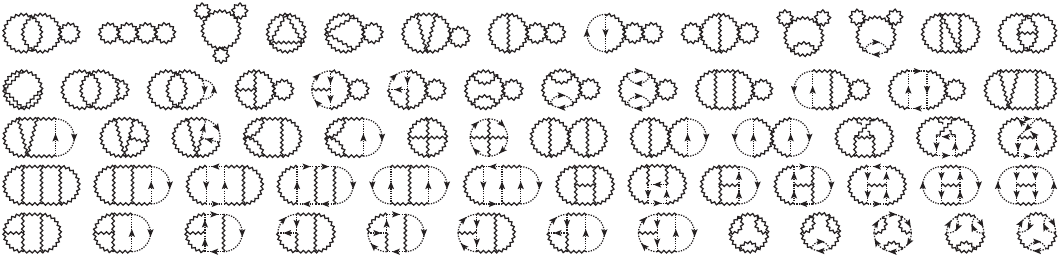}
\caption{All 65 four-loop 1PI vacuum diagrams of $\mathrm{SU}(N_c)$ Yang-Mills theory contributing to its pressure, with wiggly (dotted) lines denoting gluons (ghosts).
We note that all nonplanar graphs appearing here vanish identically due to their color factors.}
\label{fig:graphs}
\end{figure*}

The weak-coupling expansions of the pressures of both pure Yang-Mills theory and QCD reflect this separation of scales.
The leading $g^0$ and $g^2$ terms respectively arise from one- and two-loop vacuum diagrams in the full theory \cite{Shuryak:1977ut}, while soft electric fluctuations first contribute at order $g^3$ \cite{Kapusta:1979fh} and also generate a $g^4\ln g$ term at the next order, determined by Toimela in 1982 \cite{Toimela:1982hv}.
The full $g^4$ term, corresponding to three-loop full-theory and two-loop EQCD diagrams, was completed by Arnold and Zhai only considerably later in 1994, reflecting the technical challenges associated with evaluating three-loop sum-integrals \cite{Arnold:1994ps}.

The $g^5$ term, which receives contributions from three-loop soft diagrams and corrections to the EQCD matching parameters, was obtained shortly thereafter by Zhai and Kastening \cite{Zhai:1995ac} and independently by Braaten and Nieto \cite{Braaten:1995jr}.
The latter article also cast the scale separation into a systematic EFT framework, introduced the terminology EQCD and MQCD, and even outlined parts of the $g^6$ calculation.
In 2003, the QCD results were finally generalized to nonzero quark chemical potentials, allowing the extraction of various quark number susceptibilities \cite{Vuorinen:2002ue,Vuorinen:2003fs}.

At order $g^6$, contributions from the hard and soft sectors are joined by the leading ultrasoft contribution from the magnetostatic fields $A_i$.
As Linde demonstrated \cite{Linde:1980ts} (see also \cite{Gross:1980br}), this contribution is intrinsically nonperturbative, implying that it cannot be captured by a simple perturbative expansion of the path integral for the partition function.
Dimensional reduction nevertheless isolates the magnetic physics from the perturbatively calculable terms, allowing the $g^6\ln\,g$ term in the pressure to be extracted within the EFT framework --- a task accomplished by Kajantie et al.~in 2003 \cite{Kajantie:2002wa}.

The nonperturbative coefficient entering the $O(g^6)$ term was subsequently determined for $\mathrm{SU}(3)$ using three-dimensional lattice simulations and lattice-to-$\overline{\mathrm{MS}}$ matching in \cite{Hietanen:2004ew,DiRenzo:2006nh}. 
While the lattice measurements were later extended to other values of $N_c$ \cite{Hietanen:2006rc}, the corresponding four-loop matching remains incomplete for general $N_c$ \cite{Schroder:2006vz}.
For $\mathrm{SU}(3)$, the only ingredient still missing from the complete $O(g^6)$ pressure is thus the nonlogarithmic hard-scale contribution, obtained from the sum of all one-particle-irreducible (1PI)  four-loop vacuum graphs of the full theory.
Long considered intractable, this problem has remained open for over two decades.

In this Letter, we determine this missing hard-scale contribution to the $O(g^6)$ pressure of pure $\mathrm{SU}(N_c)$ Yang-Mills theory.
Building on the reduction of four-loop vacuum diagrams to canonical sum-integrals \cite{Navarrete:2024ruu}, we evaluate the altogether nine unknown four-loop sum-integrals using the thermal loop-tree duality (LTD) method.
Originally developed for multiloop Feynman integrals in collider physics \cite{Catani:2008xa,Bierenbaum:2010cy,Capatti:2019ypt,Runkel:2019yrs,Capatti:2020ytd,JesusAguilera-Verdugo:2020fsn,Aguilera-Verdugo:2020set,Capatti:2022mly}, LTD was recently extended to nonzero chemical potentials \cite{Navarrete:2024zgz,Karkkainen:2025nkz,Karkkainen:2026yak} and temperature \cite{Navarrete:2025yxy,Gorda:2025cwu}, allowing thermal sum-integrals to be numerically evaluated in momentum space.
Here, we combine this framework with a unified local subtraction procedure for ultraviolet (UV) and infrared (IR) divergences, including novel IR counterterms, to complete the four-loop calculation.
Combining our result with the known contributions from EQCD and MQCD, we demonstrate the analytic cancellation of IR divergences between the different terms, allowing us to write down the complete $O(g^6)$ pressure of hot $\mathrm{SU}(3)$ Yang-Mills theory.
This marks the first order at which all three momentum scales contribute to the weak-coupling expansion, potentially improving its predictive power \cite{Laine:2003ay}.

Finally, we note that the pressure of pure $\mathrm{SU}(3)$ Yang-Mills theory has also been repeatedly determined using high-precision lattice simulations.
The classic results of Ref.~\cite{Boyd:1996bx} were subsequently extended to temperatures as high as $1000$ times the deconfinement temperature $T_c\approx$ 260 MeV \cite{Borsanyi:2012ve}, while an independent determination using shifted temporal boundary conditions has reached $230T_c$ with approximately $0.5\%$ precision \cite{Giusti:2016iqr}.
Such lattice calculations have even enabled indirect estimates of the previously unknown $g^6$ coefficient \cite{Schroder:2006vz}, providing a quantitative benchmark for our result.

\section{The four-loop pressure}

We write the bare hard-scale contribution to the pressure of pure SU($N_c$) Yang–Mills theory in the form
\begin{eqnarray}
p_\text{hard}^\text{bare}
= p_{\mathrm{SB}} + g_\mathrm{B}^2 N_c\, p_2 + g_\mathrm{B}^4 N_c^2\, p_3
+ g_\mathrm{B}^6 N_c^3\, p_4 + O(g_\mathrm{B}^8), \label{eq:hardp}
\end{eqnarray}
where $p_{\rm SB}=d_A\pi^2T^4/45$ denotes the Stefan–Boltzmann pressure of free gluons, $d_A=N_c^2-1$ is the dimension of the adjoint representation of SU$(N_c)$, and $g_\mathrm{B}$ stands for the bare (unrenormalized) coupling. 
Our focus in this work is the four-loop coefficient $p_4$, which enters at order $g_\mathrm{B}^6$ and receives contributions from a total of 65 four-loop vacuum diagrams involving gluons and ghosts, shown in Fig.~\ref{fig:graphs} \cite{Kajantie:2001hv}.
Carrying out the color and Lorentz algebra in a general covariant ($R_\xi$) gauge using \texttt{FORM} \cite{Davies:2026cci} produces some $10^5$ sum-integrals, which can be reduced to 21 canonical sum-integral structures by exploiting simple linear changes of loop-momentum variables, detailed in \cite{Navarrete:2024ruu}.
At this point, all dependence on the gauge parameter $\xi$ cancels, which serves as a highly nontrivial consistency check of the calculation.

Working in $d=3-2\epsilon$ spatial dimensions, $p_4$ can be written in the form
\begin{equation}
 p_4(d,T)=-\frac{d_A}{72}\sum_{n=1}^{21}c_n(d)\,I_n(d,T),
 \label{eq:canonical}
\end{equation}
where  the coefficients $c_n(d)$ are polynomials in $d$ up to degree 4 and are listed in \cite{Navarrete:2024ruu}.
As discussed in this reference, eleven of the canonical sum-integrals $I_n$ factorize into lower-loop structures, which can be mapped to known results using tensor decomposition and integration-by-parts identities. 
The remaining ten, $I_6$, $I_8$, and $I_{14}$--$I_{21}$, are genuine four-loop sum-integrals, of which the nontrivial part of $I_6$ is known from an earlier computation within scalar $\phi^4$ theory \cite{Gynther:2007bw}. 
This leaves us with nine sum-integrals, the Laurent expansions of which (up to order $\epsilon^0$) complete the determination of $p_4$.

\begin{figure}[!t]
\centering
\includegraphics[width=\columnwidth]{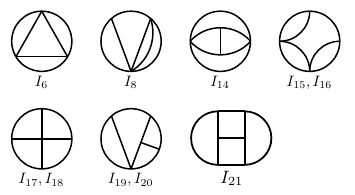}
\caption{List of the nonfactorizing four-loop vacuum topologies appearing in the canonical pressure of Yang-Mills theory \cite{Navarrete:2024ruu} along with the sum-integrals associated with each topology.}
\label{fig:sectors}
\end{figure}

The canonical sum-integrals all have the generic form
\begin{eqnarray}
 I_n&=&{}\sumint_{K_1}\sumint_{K_2}
 \sumint_{K_3}\sumint_{K_4}\,
 \mathcal I_n(K_i^2;K_i\cdot K_j),\label{eq:family}\\
 \sumint_{K_i}&\equiv{}&\bigg(\frac{e^{\gamma_\text{E} }\bar{\Lambda}^2}{4\pi}\bigg)^\epsilon T\sum_{n_i\in\mathbb Z}
 \int\!\frac{\mathrm{d}^d \mathbf{k}_i}{(2\pi)^d},
 \label{eq:measure}
\end{eqnarray}
where $\bar{\Lambda}$ is the \MSbar~scale and $K_i=(2\pi n_iT, \mathbf{k}_i)$, $i=1,\ldots,4$, denote independent bosonic loop momenta in Euclidean signature, with $n_i\in\mathbb Z$ labeling the Matsubara frequencies. 
The integrands $\mathcal I_n$ are on the other hand rational functions of the squared loop momenta $K_i^2$ and their scalar products $K_i\cdot K_j$ ($i<j$) so that, for example, the integral $I_{16}$ reads
\begin{equation}
\begin{split}
I_{16}= &\sumint_{K_1}\sumint_{K_2}
             \sumint_{K_3}\sumint_{K_4} \frac{\big[(K_1-K_2-K_3)^2\big]^3}
{K_1^2K_2^2K_3^2[K_4^2]^3} \\
 & \times\frac{1}{(K_1-K_4)^2(K_2-K_4)^2(K_3-K_4)^2}.
 \label{eq:examplemaster}
\end{split}
\end{equation}
Its high propagator power and nontrivial numerator illustrate the complicated ultraviolet and infrared sensitive structures encountered in the calculation.
All ten nonfactorizing integrals fall into the seven vacuum integral topologies shown in Fig.~\ref{fig:sectors}, while their canonical index definitions are collected in the End Matter.

\section{Evaluation of the sum-integrals}

We evaluate all ten nonfactorizing sum-integrals, including the previously known $I_6$, using a unified subtraction procedure that removes both UV and IR divergences at the integrand level.
Here, we follow the local subtraction approach of Refs.~\cite{Capatti:2019edf,Capatti:2022tit} and extend it to thermal soft regions.
Building on the Bogoliubov $R$-operation~\cite{Bogoliubov:1957gp,Hepp:1966eg,Zimmermann:1969jj}, this procedure systematically accounts for nested and overlapping divergences, including mixed UV-IR regions. 
The IR counterterms are constructed by identifying soft subgraphs associated with vanishing Matsubara frequencies and performing the corresponding soft-momentum expansions.

Within thermal LTD, the Matsubara sums are always performed analytically, yielding closed-form representations in terms of three-dimensional spatial momenta \cite{Navarrete:2025yxy}.
After the local subtractions, the resulting integrands are both UV and IR finite so that they can be evaluated numerically using Monte-Carlo integration with \texttt{Symbolica}~\cite{Ruijl:Symbolica:2026}, requiring an estimated $10^{6}$ CPU hours of supercomputing time.
Schematically, each sum-integral is decomposed in the form $I_n = I_n^{\mathrm{fin}} + I_n^{\mathrm{CT}}$, where $I_n^{\mathrm{CT}}$ collects all subtraction counterterms and $I_n^{\mathrm{fin}}$ denotes the finite and numerically computable remainder.
The counterterms $I_n^{\mathrm{CT}}$ are then evaluated analytically in dimensional regularization using \texttt{FIRE}~\cite{Smirnov_2024} and \texttt{Reduze2}~\cite{vonmanteuffel2012reduze2distributed} to reduce them to master integrals, the expansions of which are taken from the literature~\cite{schroder2003highprecisionevaluationfourloopvacuum,Rajantie_1996,Kajantie_2003,Schr_der_2005,Schr_der_2026,Arnold:1994ps,Ghisoiu:2012yk}.

\begin{figure*}[t!]
\centering
\includegraphics[width=0.48\textwidth]{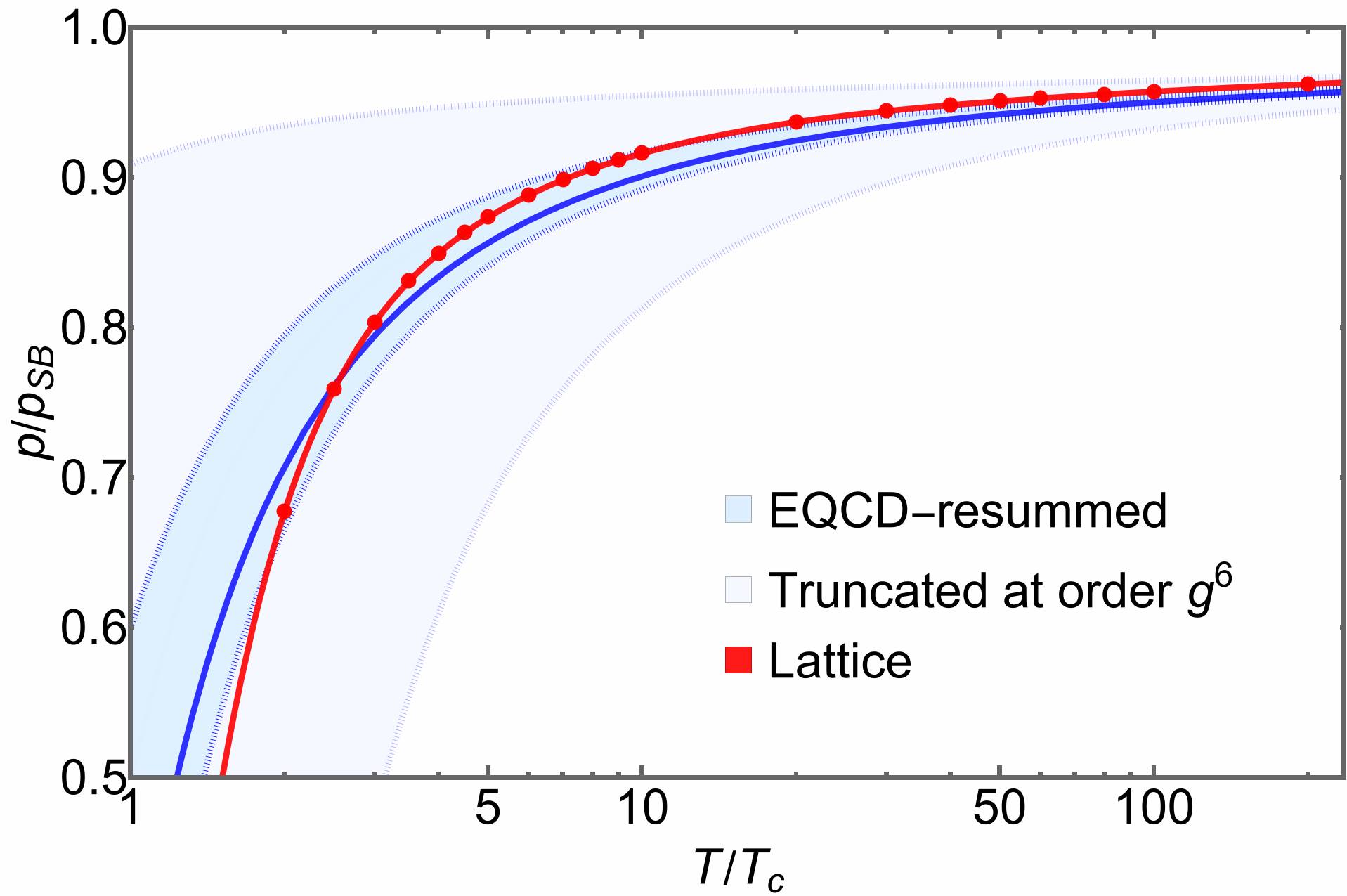}\hfill
\includegraphics[width=0.48\textwidth]{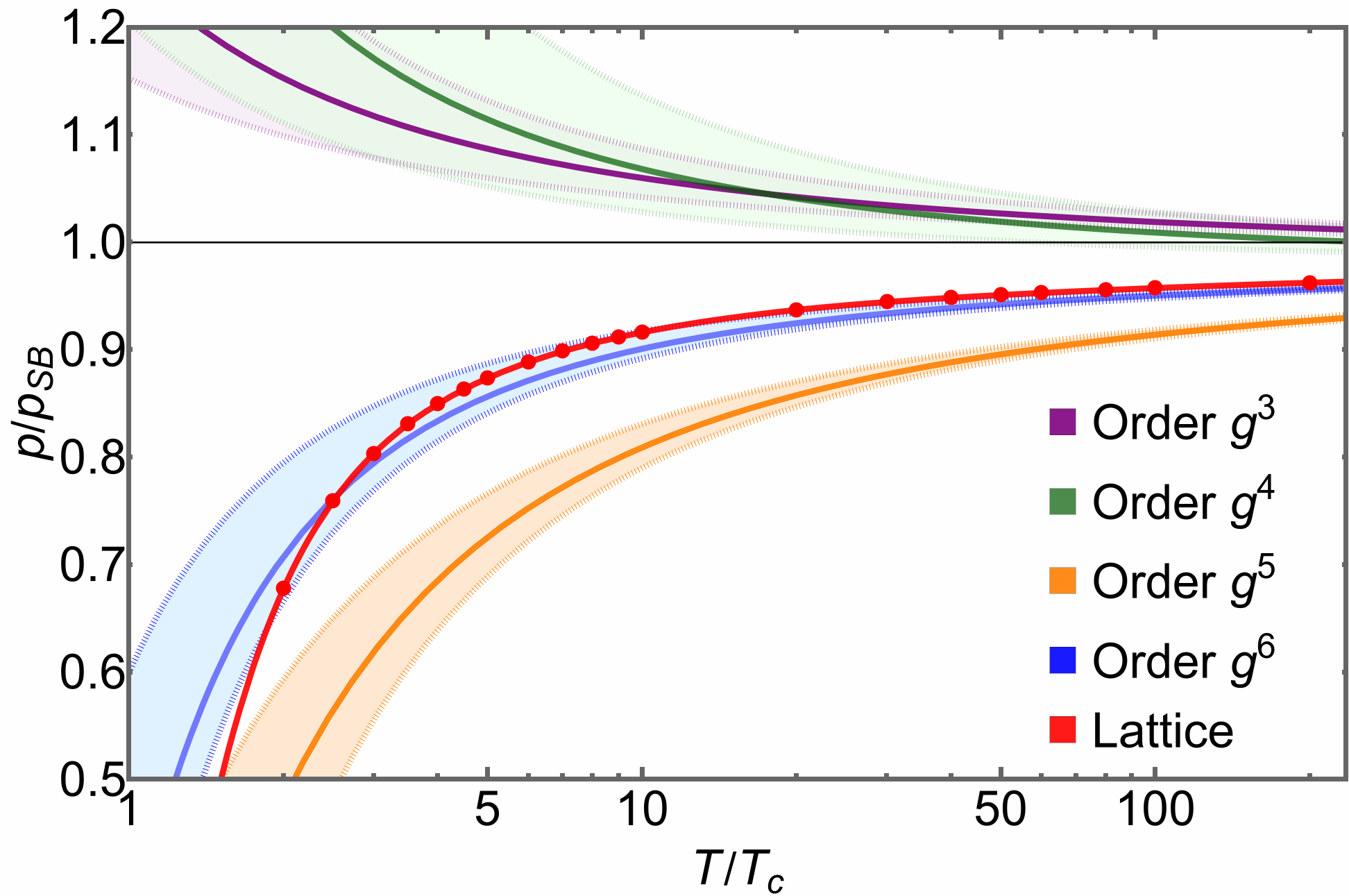}
\caption{Left: comparison of the $O(g^6)$ SU(3) Yang-Mills pressure (blue bands) with four-dimensional lattice results (red data points; negligible errors) from Ref.~\cite{Borsanyi:2012ve}, with both results normalized by the noninteracting Stefan-Boltzmann limit $p_\text{SB}$.
The shaded bands indicate the combined uncertainties from renormalization-scale variation and numerical uncertainties in the hard and magnetic $O(g^6)$ contributions. 
To convert the perturbative result to the units quoted here, we use the three-loop running coupling and the value $T_c/\Lambda_{\overline{\mathrm{MS}}}\approx 1.26$ reported in Ref.~\cite{Borsanyi:2012ve}. \\
Right: comparison of the four highest perturbative orders available for the normalized pressure of SU(3) YM theory: $O(g^3)$, $O(g^4)$, $O(g^5)$, and $O(g^6)$.
All results use the same EQCD-resummation prescription, implemented at the respective perturbative orders, with scale variation following that of the left panel.}
\label{fig:result}
\end{figure*}

Combining the numerical remainders with the analytic counterterms yields bare results for all ten nonfactorizing sum-integrals, including an independent validation of the $I_6$ result from \cite{Gynther:2007bw}.
The analytically determined coefficients of the $1/\epsilon^n$ poles further allow us to verify the cancellation of UV divergences upon renormalization of the bare coupling $g_\mathrm{B}$ and of the remaining IR divergences against EQCD and MQCD contributions. 
The explicit Laurent expansions for the $I_n$ are finally listed in the End Matter.

\section{Results}

Combining our results for the new nonfactorizing sum-integrals with the factorizing contributions in Eq.~\eqref{eq:canonical}, we may now finally write down the $O(g^6)$ hard-scale contribution to the Yang-Mills pressure, $p_{\rm E}$. 
Following Ref.~\cite{Schroder:2006vz}, we write $\hat g^2\equiv N_c g^2(\bar{\Lambda})/(16\pi^2)$ and $L\equiv\ln[\bar{\Lambda}/(4\pi T)]$, where $g(\bar{\Lambda})$ is the renormalized coupling related to the bare coupling of Eq.~\eqref{eq:hardp} through
\begin{eqnarray}
\frac{N_c g_B^2}{16\pi^2}&=&\hat g^2\bigg[1-\frac{11}{3\epsilon}\hat g^2+\bigg(\frac{121}{9\epsilon^2}-\frac{17}{3\epsilon}\bigg)\,\hat g^4+O(\hat g^6)\bigg]\, .
\end{eqnarray}
The renormalized pressure then becomes
\begin{equation}
\left.\frac{p_{\rm E}}{p_{\rm SB}}\right|_{\hat g^6}
=\hat g^6\bigg[
\frac{c_{-1}(L)}{\epsilon}
+c_2 L^2+c_1 L+\tilde\beta_{\rm E1}
\bigg]\, ,
\label{eq:hardresult}
\end{equation}
where the pole and logarithmic coefficients $c_{-1}$, $c_2$, and $c_1$ are determined analytically and agree with the predictions in Eqs.~(2.34)-(2.36) of Ref.~\cite{Schroder:2006vz}.
In particular, the hard-scale IR poles were previously inferred by requiring their cancellation against the ultraviolet divergences of EQCD and MQCD, but our explicit four-loop calculation has now independently confirmed this.

The main result of our present work is the value of the previously unknown four-loop coefficient
\begin{equation}
\tilde\beta_{\rm E1}=6236\pm40,
\label{eq:betaE1}
\end{equation}
where the uncertainty originates from the numerical evaluation of the sum-integrals.
This agrees with the earlier lattice-based estimate $\tilde\beta_{\rm E1}\simeq6000\pm600$ \cite{Schroder:2006vz}, but is now obtained directly through an explicit four-loop calculation.
Following the notation of Kajantie et al.~\cite{Kajantie:2002wa} in their Eq.~(6.11), our result translates to the nonlogarithmic order-$g^6$ coefficient
\begin{equation}
q_c(N_c=3,N_f=0)=-3591.2\pm16.9_{\rm pert}
\pm250_{\rm mag},
\label{eq:qc}
\end{equation}
where the first uncertainty originates from our result and the second reflects the uncertainty in the nonperturbative magnetic constant $B_{\rm G}$ in \cite{DiRenzo:2006nh}.

To inspect the behavior of the $O(g^6)$ pressure, in Fig.~\ref{fig:result} we compare our new perturbative expression to the four-dimensional lattice result of~\cite{Borsanyi:2012ve}. 
The two perturbative bands in the left panel correspond to a strict expansion in $g$, truncated at order $g^6$ (light blue band), and its EQCD-resummed counterpart  \cite{Blaizot:2003iq,Laine:2006cp}, where the effective-theory parameters are kept unexpanded (darker blue band) 
\footnote{More specifically, this amounts to using two-loop results for $m_E^2$ and $g_E^2$ within the (analytic) EQCD pressure and not expanding the result in powers of the full-theory coupling. 
Following Ref.~\cite{Laine:2006cp}, we also re-express the hard contribution in terms of $g_E^2$, including compensating matching terms to preserve the strict expansion up to and including $O(g^6)$.}.
In both cases, the bands are generated by varying the \MSbar~renormalization scale $\bar{\Lambda}$ by a factor of two around the central value $\bar{\Lambda}=4\pi e^{-\gamma_\text{E}-1/22}T$, obtained by applying the Fastest Apparent Convergence (FAC) criterion to the EQCD gauge coupling $g_\text{E}$ \cite{Laine:2006cp}.

As expected, we observe a marked decrease of the renormalization scale dependence upon the EQCD resummation of \cite{Blaizot:2003iq,Laine:2006cp}.
Remarkably, the resummed band is seen to be in good quantitative agreement with the lattice data all the way down to temperatures a few times the critical temperature of the deconfinement transition, reported as $T_c \approx 260$ MeV in \cite{Borsanyi:2012ve}.

The favorable behavior of the $O(g^6)$ result is in striking contrast to that of lower perturbative orders, inspected in the right panel of Fig.~\ref{fig:result}. 
In agreement with earlier literature (see, e.g., \cite{Blaizot:2003iq}), we observe the $O(g^3)$ and $O(g^4)$ calculations predicting pressure values exceeding the free Stefan-Boltzmann limit, while the sizable and negative $g^5$ correction brings the pressure well below the lattice results (note, however, that other resummation schemes have reported improved behavior already at lower loop orders \cite{Blaizot:2003iq,Andersen:2010ct}).
The fact that, within this resummation scheme, quantitative agreement with the lattice data is achieved only at order $g^6$ provides strong support for the `Laine conjecture' that the weak-coupling expansion should become quantitatively reliable once contributions from all three momentum scales are included \cite{Laine:2003ay}.

\section{Discussion}

In this Letter, we have completed a task set out over two decades ago: determining the full order-$g^6$ pressure of hot SU(3) Yang-Mills theory.
Combined with known chromoelectric and chromomagnetic contributions, our evaluation of the 65 four-loop vacuum diagrams for general $N_c$ yields the full weak-coupling expansion up to this order for SU(3), marking the second time the thermodynamics of a high-temperature quantum field theory have been evaluated at this order \cite{Gynther:2007bw} and the first time this has been achieved in a physical gauge theory \footnote{See, however, also the recent determination of the four-loop pressure of cold and dense QED \cite{Karkkainen:2026yak}.}.

Beyond determining the previously unknown hard-scale coefficient, our calculation demonstrates the analytic cancellation of infrared divergences against the corresponding effective-theory contributions.
This provides a nontrivial verification of the dimensional reduction paradigm at the first order where all three momentum scales contribute, including the intrinsically nonperturbative magnetic sector.

With the $O(g^6)$ pressure now fully determined, we can directly compare our result with data from four-dimensional lattice calculations \cite{Borsanyi:2012ve,Giusti:2016iqr}.
To facilitate this, we suppress the intrinsically sizable renormalization-scale dependence of the perturbative pressure by performing the so-called EQCD resummation, first suggested in \cite{Blaizot:2003iq} and further developed in \cite{Laine:2006cp}.
Doing so, we find that while previous perturbative orders fail to reproduce the behavior of lattice data, our new result is in  quantitative agreement with the data of Ref.~\cite{Borsanyi:2012ve}  over a broad range of temperatures.

Finally, we note that the two key ingredients enabling the completion of this computation were the preparatory work performed in \cite{Navarrete:2024ruu}, reducing the problem to the evaluation of a handful of four-loop sum-integrals, and the development of the thermal loop-tree duality method \cite{Navarrete:2025yxy} that enabled the efficient numerical handling of these integrals when combined with proper local ultraviolet and infrared subtractions.
The natural next steps include determining the order-$g^7$ contribution, which requires five-loop vacuum integrals in three-dimensional EQCD, as well as extending the calculation to full QCD with dynamical quarks and nonzero chemical potentials. The latter will eventually enable us to explore regimes beyond the reach of modern lattice calculations and relevant for neutron-star mergers. With thermal LTD now benchmarked at four-loop order both at high temperatures and high chemical potentials \cite{Navarrete:2024zgz,Karkkainen:2025nkz,Karkkainen:2026yak}, no conceptual obstacles prevent progress toward this extension.

\emph{Acknowledgements.}---%
While finalizing this work, we became aware of an independent effort by a research group at the University of Granada \cite{Chala:g6}, aimed at completing the same calculation. 
Prior to coordinated arXiv submission, we verified agreement between our main results, but have not otherwise read the article of the other collaboration.

We thank Zeno Capatti, Valentin Hirschi, and Aleksi Kurkela for useful discussions.
The work has been supported by the Research Council of Finland through projects 354533 (AV) and 347499 (RP) as well as the Centre of Excellence in Neutron-Star Physics, project 374062 (AV); the European Research Council project ExPertQCD, grant No.~101231521 (RP); and ANID under FONDECYT project No.~1231056 and Exploraci\' on Project No.~13250014 (YS). Computational resources have been provided by CSC - IT Center for Science, Finland.

\emph{AI statement.}---%
ChatGPT (OpenAI) was used for stylistic editing and to improve the presentation and discussion of the results, but no AI tools were used to perform or validate the calculations.

\bibliography{refPRL.bib}

\clearpage

\section*{End matter}

\appendix

Here we specify the canonical definitions and Laurent
expansions of the ten nonfactorizing four-loop sum-integrals entering the Yang-Mills pressure. 
Further details of their evaluation, including the unified UV and IR subtraction algorithms and explicit four-loop integral evaluations, will be presented in a forthcoming paper.

Following \cite{Navarrete:2024ruu}, we introduce the sum-integral family
\begin{equation}
I(s_1,\ldots,s_{10})
\equiv
\sumint_{K_1}\sumint_{K_2}
\sumint_{K_3}\sumint_{K_4}
\prod_{a=1}^{10}\frac{1}{[P_a^2]^{s_a}},
\label{eq:canonicalfamily}
\end{equation}
where $s_a\in\mathbb Z$ and the propagator momenta
are defined by
\begin{align}
&\{P_1,\ldots,P_{10}\}=
\{K_1,K_2,K_3,K_4,K_1-K_4,K_2-K_4,\nonumber\\
&\quad K_3-K_4,K_1-K_2,
K_1-K_3,K_1-K_2-K_3\}.
\label{eq:momentumfamily}
\end{align}
The indices $s_a$ specify propagator powers: positive values correspond to denominator factors, zero to an absent propagator, and negative values to numerator factors, with Euclidean momenta satisfying $P_a^2=p_{a,0}^2+\mathbf{p}_a^2$.
Table~\ref{tab:masters} lists the canonical index tuples $(s_1,\ldots,s_{10})$ for the ten nonfactorizing integrals $I_6$, $I_8$, and $I_{14}$--$I_{21}$.

We immediately notice that all ten integrals satisfy $\sum_{a=1}^{10}s_a=6$ and, including
the \MSbar~factors in the measure of Eq.~\eqref{eq:measure}, have mass dimension $4$ in $d=3-2\epsilon$ spatial dimensions. 
Their Laurent expansions can thus be written in the form
\begin{equation}
I_n=\frac{T^{4-8\epsilon} {\bar{\Lambda}}^{8\epsilon}}{16(4\pi)^{4+8\epsilon}}
\left[
\frac{A_n}{\epsilon^2}
+\frac{B_n}{\epsilon}
+C_n+O(\epsilon)
\right],
\label{eq:masterlaurent}
\end{equation}
where $\bar{\Lambda}$ again denotes the $\overline{\mathrm{MS}}$ renormalization scale.
The analytically evaluated counterterms $I_n^{\mathrm{CT}}$ determine the pole coefficients $A_n$ and $B_n$ and contribute to the finite coefficients $C_n$, which also receive the numerically evaluated remainders $I_n^{\mathrm{fin}}$.
The coefficients of all ten integrals are collected in Table~\ref{tab:masters}.

The counterterms $I_n^{\mathrm{CT}}$ include the $O(\epsilon)$ coefficients $b$ and $s$ of the three-loop basketball $B$ and spectacle $S$ sum-integrals, defined in Eq.~\eqref{eq:ThreeLoopIntegrals} below. 
While these two integrals are only known up to their constant parts \cite{Arnold:1994ps,Braaten:1995jr,Schroder:2012hm,Nishimura:2012ee}, it turns out to be sufficient to keep $b$ and $s$ in symbolic form, as they cancel in the full pressure after accounting for the respective terms generated by the renormalization of the 3-loop hard pressure $p_3$, to which both $B$ and $S$ contribute \cite{Nishimura:2012ee,Navarrete:2024ruu}. 
For reference, we note that in the notation of Ref.~\cite{Nishimura:2012ee}, canonicalization gives $S = B_3 / 4- B_1 /2$, where $B_1$ is known analytically and $B_3$ can be obtained from $S$ by replacing its numerator with  $[(K_2-K_3)^2]^2$. 
The latter is known up to its constant term, while its $O(\epsilon)$ coefficient contains $s$.

\begin{widetext}
\begin{align}\label{eq:ThreeLoopIntegrals}
    B &\equiv \sumint_{K_1}\sumint_{K_2}\sumint_{K_3} \frac{1}{K_1^2K_2^2(K_1-K_3)^2(K_2-K_3)^2}
 \;=\;\frac{T^{4-6\epsilon}{\bar{\Lambda}}^{6\epsilon}}{(4\pi)^{2+6\epsilon}}\Big[ \frac{1}{24 \epsilon }+\frac{1}{24} \Big(\frac{91}{15}+8Z_1-2Z_3\Big)+b\, \epsilon   +\mathcal{O}(\epsilon^2)\Big] \,,\\ \nonumber
    S &\equiv \sumint_{K_1}\sumint_{K_2}\sumint_{K_3} \frac{[K_2\cdot K_3]^2}{[K_1^2]^2K_2^2K_3^2(K_1-K_2)^2(K_1-K_3)^2} 
    \;=\;\frac{T^{4-6\epsilon}{\bar{\Lambda}}^{6\epsilon}}{(4\pi)^{2+6\epsilon}}\Big[
   \frac{1}{108 \epsilon }+\frac{49+12\gamma_\text{E}+104Z_1-20Z_3}{1728}+s\, \epsilon +\mathcal{O}(\epsilon^2)
    \Big] \,.
\end{align}

\refstepcounter{table}\label{tab:masters}

\begin{center}
\scriptsize
\begin{ruledtabular}
\begin{tabular}{clccc}
$I_n$ & $(s_1,\ldots,s_{10})$ & $A_n$ & $B_n$ & $C_n$ \\
\hline
\vspace{0.5mm}
$I_6$    & $(1,1,1,0,1,1,1,0,0,0)$    & $1$    & $12 Z_1-4 Z_3+\frac{44}{5}  -\frac{\pi ^2}{2}$ & $(-130.278\pm 0.004)+48 b$ \\
\vspace{0.5mm}
$I_8$    & $(1,1,1,1,1,0,0,0,0,1)$    & $1/3$  & $4 Z_1-\frac{4 Z_3}{3}+\frac{71}{60}$ & $(-45.486\pm 0.008)+16 b$ \\
\vspace{0.5mm}
$I_{14}$ & $(1,1,1,1,1,1,0,0,-1,1)$   & $1/12$ & $\frac{11 Z_1}{3}-\frac{5 Z_3}{3}+\frac{37}{72}-\frac{4 \gamma_\text{E} }{3}$ & $(-94.420\pm 0.012)+20 b$ \\
\vspace{0.5mm}
$I_{15}$ & $(1,1,1,2,1,1,1,0,0,-2)$   & $37/18$ & $14 Z_1-\frac{26 Z_3}{9}+\frac{1961}{270}+\frac{16 \gamma_\text{E} }{3}-\frac{5 \pi ^2}{16}$ & $(6.84\pm 0.09)+8 b+192 s$ \\
\vspace{0.5mm}
$I_{16}$ & $(1,1,1,3,1,1,1,0,0,-3)$   & $-499/180$ & $-19 Z_1+\frac{178 Z_3}{45}-\frac{157}{36}+\frac{13}{225} \zeta (3)-\frac{107 \gamma_\text{E} }{15}+\frac{7 \pi ^2}{16}$ & $(123.7\pm 3.8)+12 b-480 s$ \\
\vspace{0.5mm}
$I_{17}$ & $(1,1,1,1,1,1,-1,-1,1,1)$ & $-5/12$ & $-\frac{7 Z_1}{3}+\frac{Z_3}{3}-\frac{5389}{360}-\frac{4 \gamma_\text{E} }{3}+\frac{4 \pi ^2}{3}$ & $(-36.865\pm 0.022)-4 b$ \\
\vspace{0.5mm}
$I_{18}$ & $(1,1,1,1,1,1,0,-2,1,1)$  & $1/6$ & $\frac{22 Z_1}{3}-\frac{10 Z_3}{3}+\frac{31}{36}-\frac{8 \gamma_\text{E} }{3}$ & $(-191.151\pm 0.028)+40 b$ \\
\vspace{0.5mm}
$I_{19}$ & $(1,1,1,1,1,1,1,1,-2,0)$  & $-5/3$ & $-\frac{32 Z_1}{3}+2 Z_3-\frac{937}{180}-\frac{14 \gamma_\text{E} }{3}$ & $(-69.752\pm 0.012)-24 b$ \\
\vspace{0.5mm}
$I_{20}$ & $(1,1,1,1,1,1,1,1,0,-2)$  & $-43/18$ & $-\frac{38 Z_1}{3}+\frac{14 Z_3}{9}-\frac{3229}{270}-8 \gamma_\text{E} +\frac{\pi ^2}{2}$ & $(-181.273\pm 0.011)-\frac{56 b}{3}$ \\
\vspace{0.5mm}
$I_{21}$ & $(1,1,1,1,1,1,1,1,1,-3)$  & $-16/3$ & $-24 Z_1+\frac{4 Z_3}{3}-\frac{2972}{45}-20 \gamma_\text{E} +\frac{11 \pi ^2}{2}$ & $(-501.63\pm 0.11)-16 b$
\end{tabular}
\end{ruledtabular}
\end{center}
\noindent\textbf{TABLE~\thetable.}
The ten nonfactorizing four-loop sum-integrals in the canonical notation of Ref.~\cite{Navarrete:2024ruu}, with the integer tuple defining $I_n=I(s_1,\ldots,s_{10})$. 
The Laurent coefficients follow Eq.~\eqref{eq:masterlaurent}, where $\gamma_\text{E} = 0.5772...$ denotes the Euler-Mascheroni constant and $Z_n\equiv \zeta'(-n)/\zeta(-n)$ with Riemann zeta function $\zeta$.
For comparison, the semi-analytic result for $I_6$ from Ref.~\cite{Gynther:2007bw} reads $C_6=(-130.27732347(65\pm20))+48b$.
\end{widetext}

\end{document}